\documentclass[preprint,showpacs,preprintnumbers,amsmath,amssymb,showkeys]{revtex4}

\usepackage{graphicx}
\usepackage{dcolumn}
\usepackage{bm}

\begin{document}

\preprint{}

\title{Omnidirectional total transmission at the interface associated with
an anisotropic dielectric-magnetic metamaterial}

\author{Hailu Luo} \thanks{Author to whom correspondence should be addressed.
E-mail: hailuluo@sohu.com}
\author{Weixing Shu}
\author{Fei Li}
\author{Zhongzhou Ren}
\affiliation{Department of Physics, Nanjing University, Nanjing
210008, China}
\date{\today}

\begin{abstract}
Based on the Ewald-Oseen extinction theorem, the omnidirectional
total transmission of waves incident from vacuum into an
anisotropic dielectric-magnetic metamaterial is investigated. It
is shown that the omnidirectional total transmission need not
limit at the interface associated with the conventional
nonmagnetic anisotropic medium. The recent advent of a new class
of anisotropic dielectric-magnetic matermaterial make the
omnidirectional total transmission become available. It is found
that the inherent physics underlying the omnidirectional total
transmission are collective contributions of the electric and
magnetic responses.
\end{abstract}

\maketitle

\section{Introduction}\label{Introduction}
The phenomena of reflection and refraction of light at the
interface of two transparent media are widely used for steering
light in many optical devices. There has been much discussion on
realizations of omnidirectional total
reflection~\cite{Fink1998,Dowling1998,Winn1998,Deopura2001,Shandon2002,Han2005}.
While the omnidirectional total transmission, where wave is
completely transmitted for arbitrary incident directions,
attracted little attention.  Recently, the omnidirectional total
refraction at the planar interface associated with convention
nonmagnetic uniaxial crystal have been
studied~\cite{Zhang2003,Liu2004}.

The advent of a new class of anisotropic dielectric-magnetic
metamaterial with negative permittivity and permeability has
attained considerable
attention~\cite{Veselago1968,Smith2000,Shelby2001,Parazzoli2003,Houck2003}.
Lindell {\it et al.} \cite{Lindell2001} have extended that
anomalous negative refraction can occur at the interface
associated with an anisotropic dielectric-magnetic metamaterial,
which does not necessarily require that all tensor elements of
permittivity $\boldsymbol{\varepsilon}$  and permeability
$\boldsymbol{\mu}$ have negative values. The study of such
anisotropic metamaterial have recently received much interest and
attention~\cite{Hu2002,Zhou2003,Smith2003,Smith2004,Luo2005,Luo2006}.

The question thus naturally arises: whether there exists any type
of interface associated the anisotropic dielectric-magnetic
material support the omnidirectional total transmission. In the
present work, we present an investigation on the omnidirectional
total transmission from vacuum into anisotropic
dielectric-magnetic metamaterial. We find that the omnidirectional
total transmission need not limit at the interface associated with
conventional nonmagnetic uniaxial crystal. We want to explore how
the omnidirectional total transmission shows up at microscopic
leave.

\section{The Ewald-Oseen extinction theorem}\label{sec2}
In molecular optics theory, a bulk material can be regarded as a
collection of molecules (or atoms) embedded in the vacuum. Under
the action of an incident field, the molecules oscillate as
electric and magnetic dipoles and emit radiations. The radiation
field and the incident field interact to form the new transmitted
field in the material and the reflection field outside the
material~\cite{Born1999}. Let us carry this idea one step further:
the metamaterial is structured into subunits. In the case of
electromagnetic radiation this usually means that the subunits
must be much smaller than the wavelength of radiation. Then the
unit cells of metamaterials can be modelled as the molecules (or
atoms) in ordinary materials.

For anisotropic metamaterials one or both of the permittivity and
permeability are second-rank tensors. To simplify the proceeding
analysis, we assume the permittivity and permeability tensors are
simultaneously diagonalizable:
\begin{eqnarray}
\boldsymbol{\varepsilon}=\left(
\begin{array}{ccc}
\varepsilon_x  &0 &0 \\
0 & \varepsilon_y &0\\
0 &0 & \varepsilon_z
\end{array}
\right), ~~~\boldsymbol{\mu}=\left(
\begin{array}{ccc}
\mu_x &0 &0 \\
0 & \mu_y &0\\
0 &0 & \mu_z
\end{array}
\right).\label{matrix}
\end{eqnarray}
where $\varepsilon_i$ and $\mu_i$  are the permittivity and
permeability constants, respectively. We choose the $z$ axis to be
normal to the interface, the $x$ and $y$ axes locate at the plane
of the interface.

Let us consider a monochromatic electromagnetic field ${\bf E}_i =
{\bf E}_{i0}  \exp[i({\bf k}_i\cdot{\bf r}-\omega t)]$  incident
from vacuum into the anisotropic metamaterial. The reflected and
transmitted fields can be express as ${\bf E}_r = {\bf E}_{r0}
\exp[i({\bf k}_r\cdot{\bf r}-\omega t)]$ and ${\bf E}_r = {\bf
E}_{t0} \exp[i({\bf k}_t\cdot{\bf r}-\omega t)]$, respectively.
The incident angle is given by $\theta_i
=\tan^{-1}[k_{ix}/k_{iz}]$, and the refractive angle of the
transmitted wave vectors is decided by $\theta_t=\tan^{-1}
[k_{tx}/{k_{tz}}]$. For compactness, let us first explore the
E-polarized incident waves.

\begin{figure}
\includegraphics[width=12cm]{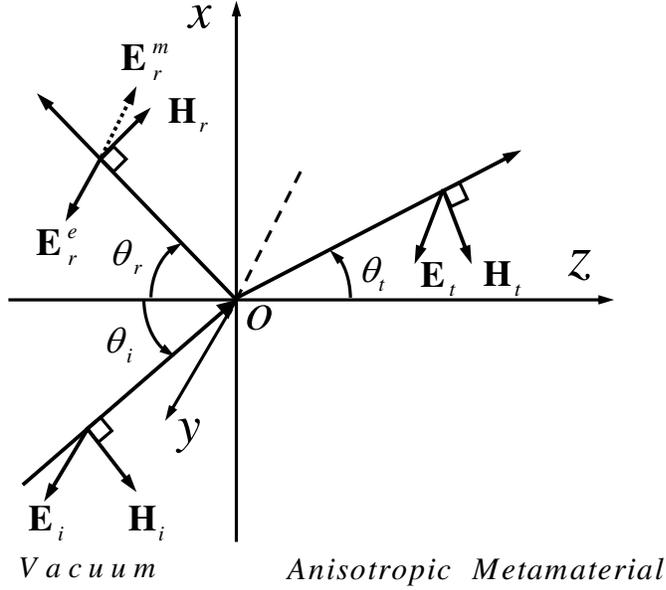}
\caption{\label{Fig1}  Schematic diagram for how the reflected and
transmitted fields of E-polarized waves are generated by the
incident field and radiated fields of dipoles. In the vacuum, the
reflected field ${\bf E}_r={\bf E}^e_{rad}+{\bf E}^m_{rad}$, while
the transmitted field ${\bf E}_t={\bf E}_i+{\bf E}^e_{rad}+{\bf
E}^m_{rad}$ in the anisotropic metamaterial.}
\end{figure}

Following the Ewald-Oseen extinction theorem, the total radiated
field ${\bf E}_{rad}$ is the sum of the contribution from all
electric dipoles ${\bf E}_{rad}^e$ and that from all magnetic
dipoles ${\bf E}_{rad}^m$. While the incident field is assumed to
permeate to the medium without being affected by the interface and
the properties of that medium~\cite{Born1999}. Hence the reflected
field in vacuum ($-\infty < z<0$) can be expressed in the terms of
the collective operations of the electric and magnetic responses:
\begin{equation}\label{Er}
{\bf E}_r={\bf E}^{e}_{rad}+{\bf E}^m_{rad}.
\end{equation}
While the transmitted field is the superstition of the incident
field and all the radiated field induced by the dipoles:
\begin{equation}\label{Et}
{\bf E}_t={\bf E}^{e}_{rad}+{\bf E}^m_{rad}+{\bf E}_i.
\end{equation}
In the other words, the re-emission of the electric and magnetic
dipoles to the half of the space ($0 \leq z<+\infty$) extinguishes
the incident field and produces the transmitted field.

The electric fields radiated by electric dipoles and magnetic
dipoles are respectively decided by \cite{Born1999}
\begin{eqnarray}\label{Eedf}
&&{\bf
E}^{e}_{rad}=\nabla(\nabla\cdot{\bf{\Pi}}_e)-\varepsilon_0\mu_0\frac{\partial^2{\bf{\Pi}}_e}{\partial
t^2},\\
&&{\bf
E}^{m}_{rad}=-\mu_0\nabla\times\frac{\partial{\bf{\Pi}}_m}{\partial
t}.\label{Emdf}
\end{eqnarray}
Here ${\bf \Pi}_e$ and ${\bf \Pi}_m$ are the Hertz vectors,
\begin{eqnarray}\label{Pie}
&&{\bf \Pi}_e({\bf r})=\int \frac{{\bf P}({\bf
r'})}{\varepsilon_0}G({\bf r}-{\bf r}')\hbox{d}{\bf r}',\\
&&{\bf \Pi}_m({\bf r})=\int {\bf M}({\bf r'})G({\bf r}-{\bf
r}')\hbox{d}{\bf r}'.\label{Pim}
\end{eqnarray}
${\bf P}$ is the dipole moment density of electric dipoles and
${\bf M}$ is that of magnetic dipoles, which are related to the
transmitted fields by ${\bf
P}=\varepsilon_0\boldsymbol{\chi}_e\cdot{\bf E}_t$ and  ${\bf
M}=\boldsymbol{\chi}_m\cdot{\bf H}_t$, where the electric
susceptibility
$\boldsymbol{\chi}_e=\boldsymbol{\varepsilon}/{\varepsilon_0}-1$
and the magnetic susceptibility
$\boldsymbol{\chi}_m=\boldsymbol{\mu}/{\mu_0}-1$. The Green
function is
\begin{equation}\label{G}
G({\bf r}-{\bf r}')=\frac{\exp{(ik_i|{\bf r}-{\bf
r}'|)}}{4\pi|{\bf r}-{\bf r}'|}.
\end{equation}
where ${\bf k}_i=k_{ix}\hat{{\bf x}}+k_{iz}\hat{{\bf z}}$ is the
incident wave vector. Inserting Eq.~(\ref{G}) into
Eqs.~(\ref{Pie}) and (\ref{Pim}), then using the delta function
definition and contour integration method, the Hertz vectors can
be evaluated as\cite{Reali1982,Fu2005}
\begin{widetext}
\begin{equation}\label{Pied}
{\bf \Pi}_e=\left\{
\begin{array}{cc}
\displaystyle  -\frac{\boldsymbol{\chi}_e\cdot{\bf
E}_{t0}\exp{(i{\bf k}_{r}\cdot{\bf r})}}{2k_{1z}(k_{1z}+k_{tz})},&-\infty<z<0\\
\displaystyle  \frac{\boldsymbol{\chi}_e\cdot{\bf
E}_{t0}\exp{(i{\bf k}_{i}\cdot{\bf
r})}}{2k_{iz}(k_{iz}-k_{tz})}+\frac{\boldsymbol{\chi}_e\cdot{\bf
E}_{t0}\exp{(i{\bf k}_t\cdot{\bf r})}}{k_{t}^2-k_{i}^2},&0\leq
z<\infty
\end{array}\right.
\end{equation}
\begin{equation}\label{Pimd}
{\bf \Pi}_m=\left\{
\begin{array}{cc}
\displaystyle  -\frac{\boldsymbol{\chi}_m\cdot{\bf
H}_{t0}\exp{(i{\bf k}_{r}\cdot{\bf r})}}
{2k_{iz}(k_{iz}+k_{tz})},&-\infty<z<0\\
\displaystyle  \frac{\boldsymbol{\chi}_m\cdot{\bf
H}_{t0}\exp{(i{\bf k}_{i}\cdot{\bf
r})}}{2k_{iz}(k_{iz}-k_{tz})}+\frac{\boldsymbol{\chi}_m\cdot{\bf
H}_{t0}\exp{(i{\bf k}_{t}\cdot{\bf r})}}{k_{t}^2-k_{i}^2},&0\leq
z<\infty
\end{array}\right.
\end{equation}
\end{widetext}
where ${\bf k}_r=k_{rx}\hat{{\bf x}}+k_{rz}\hat{{\bf
z}}$ and ${\bf k}_t=k_{tx}\hat{{\bf x}}+k_{tz}\hat{{\bf z}}$ are
the reflected and transmitted wave vectors, respectively. Here we
have used the Faraday's law ${\bf H}_t=({\bf k}_t\times{\bf
E}_t)/(\omega {\boldsymbol \mu})$ which can also be established by
the molecular theory.

The contributions from the electric and magnetic dipoles form the
reflected field ${\bf E}_{r}$. Applying Eqs.~(\ref{Pied}) and
(\ref{Pimd}) for $-\infty<z<0$ to Eqs.~(\ref{Eedf}) and
(\ref{Emdf}), we obtain
\begin{equation}\label{Ere}
{\bf E}_{rad}^e =\frac{{\bf k}_r\times[{\bf k}_r\times
(\boldsymbol{\chi}_e\cdot{\bf E}_{t0})]}{2k_{iz}(k_{iz}+k_{tz})},
\end{equation}
\begin{equation}\label{Erm}
{\bf E}_{rad}^m = \frac{{\bf
k}_r\times\{\boldsymbol{\chi}_m\cdot[{\mu_0}{\boldsymbol{\mu}}^{-1}\cdot({\bf
k}_t\times {\bf E}_{t0})]\}}{2k_{iz}(k_{iz}+k_{tz})}.
\end{equation}
where ${\bf k}_r=k_{ix}\hat{{\bf x}}-k_{iz}\hat{{\bf z}}$ is the
refractive wave vector. For E-polarized incident waves, combining
Eqs.~(\ref{Pied}) and (\ref{Pimd}) with Eq.~(\ref{Et}), we can
obtain the following dispersion relation
\begin{equation}
\frac{ k_{tx}^2}{\varepsilon_y
\mu_z}+\frac{k_{tz}^2}{\varepsilon_y \mu_x}=\omega^2.\label{D}
\end{equation}
It should be mentioned that the same dispersion relation can be
obtained from the Maxwell equations.

\begin{figure}
\includegraphics[width=12cm]{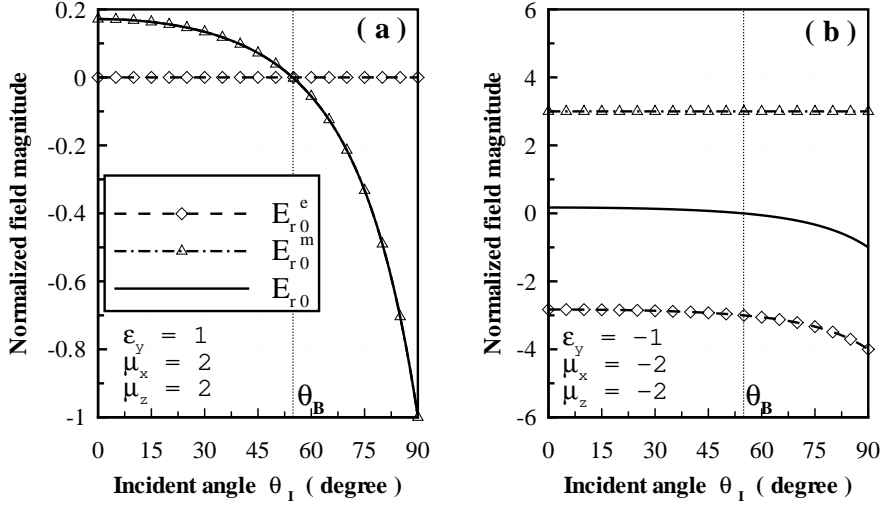}
\caption{\label{Fig2}Radiated electric field amplitudes for a
E-polarized wave incident from the vacuum into anisotropic
dielectric-magnetic material: (a) $\varepsilon_y=1$, $\mu_x=2$,
$\mu_z=2$; (b) $\varepsilon_y=-1$, $\mu_x=-2$, $\mu_z=-2$. Note
that the polarized incident wave exhibit a Brewster angle when
$E_{r0}=0$ .}
\end{figure}

Next, let us review the transmission in isotropic nonmagnetic
media. In classical electrodynamics, a simple generalization shows
that the zero flection exists occurs when H-polarized waves
incident from vacuum into an isotropic nonmagnetic medium
($\mu_I=1$). The total transmission takes place at an incident
angle satisfying $\theta_i+\theta_r=\pi/2$. Such an angle,
determined by $\theta_B=\tan^{-1}[\sqrt{\varepsilon_I}]$, is
called the Brewster angle~\cite{Born1999}.

The Ewald-Oseen extinction theorem allow us to investigate how the
Brewster angle generates. From the point of view of molecular
optics, we know that the radiated fields would never occur along
the axes of electric and magnetic dipoles. The inherent physics
for Brewster angle in isotropic nonmagnetic media is that the axes
of electric dipoles align with the direction of the reflected
wave. While in the anisotropic dielectric-magnetic material, the
physics origins are significantly different. The Brewster angle
occurs when the magnitudes of ${\bf E}_{rad}^e$ and ${\bf
E}_{rad}^m$ are equal. Setting ${\bf E}_{r}=0$ in Eq.~(\ref{Er}),
the Brewster angle can be expressed as
\begin{equation}
\theta_B=\sin^{-1}\left[\sqrt{\frac{ \mu_z(\varepsilon_y \mu_0
- \varepsilon_0 \mu_x)}{\varepsilon_0 (\mu_0^2 -\mu_x \mu_z ) }}
\right].\label{EB}
\end{equation}

For the purpose of illustration, let us choose the positive
anisotropic parameters in Fig. ~\ref{Fig2}(a), while select the
corresponding negative anisotropic parameters in Fig.
~\ref{Fig2}(b). It can be seen from the Fig.~\ref{Fig2}(a) that
${\bf E}_{rad}^e\equiv0$ when we choose
$\varepsilon_y=\varepsilon_0$. It means only the induced the
magnetic dipoles contribute to the reflected field.  In this
special case, the condition for Brewster angle requires that the
axes of magnetic dipoles align with the direction of the reflected
wave. However, compared with Fig.~\ref{Fig2}(a), the values of
${\bf E}_{rad}^e$ and ${\bf E}_{rad}^m$ in Fig.~\ref{Fig2}(b) are
much larger and never reach zero. Note that the directions of
${\bf E}_{rad}^e$ and ${\bf E}_{rad}^m$ are always reversed. When
${\bf E}^{e}_{rad}+{\bf E}^m_{rad}=0$ is satisfied, E-polarized
incident wave exhibits a Brewster angle.

\section{Omnidirectional total transmission}\label{sec3}
From the point of view of molecular optics, the microscopic
interpretation of the omnidirectional total transmission lies in
the cancellation of two extremely large fields radiated by
different types of induced dipoles:
\begin{equation}\label{Erd}
{\bf E}^{e}_{rad}+{\bf E}^m_{rad}\equiv0
\end{equation}
An alternative view is that the re-emission of the electric and
magnetic dipoles induced by incident field cancel out in vacuum.
To some extent, any incident angle can be considered to be a
Brewster angle. Substituting Eqs.~(\ref{Ere}) and (\ref{Erm}) into
Eq.~(\ref{Er}), we can obtain
\begin{equation}
\varepsilon_0 (\mu_0^2 -\mu_x \mu_z ) \sin^2 \theta_I=
\mu_z(\varepsilon_y \mu_0 - \varepsilon_0 \mu_x).\label{EO}
\end{equation}
The omnidirectional total transmission means that Eq.~(\ref{EO})
should be satisfied for arbitrary incident angle $\theta_I$. So
the conditions for omnidirectional total transmission for
E-polarized waves can be obtained as
\begin{equation}
\frac{\varepsilon_0}{\mu_0}=\frac{
\varepsilon_y}{\mu_x},~~~\mu_0^2 =\mu_x \mu_z,\label{EOC}
\end{equation}
In this case, ${\bf E}_{r0}^e\equiv0$ because  ${\bf E}_{rad}^e$
and  ${\bf E}_{rad}^m$ have the same magnitude but exhibit the
opposite signs.

\begin{figure}
\includegraphics[width=12cm]{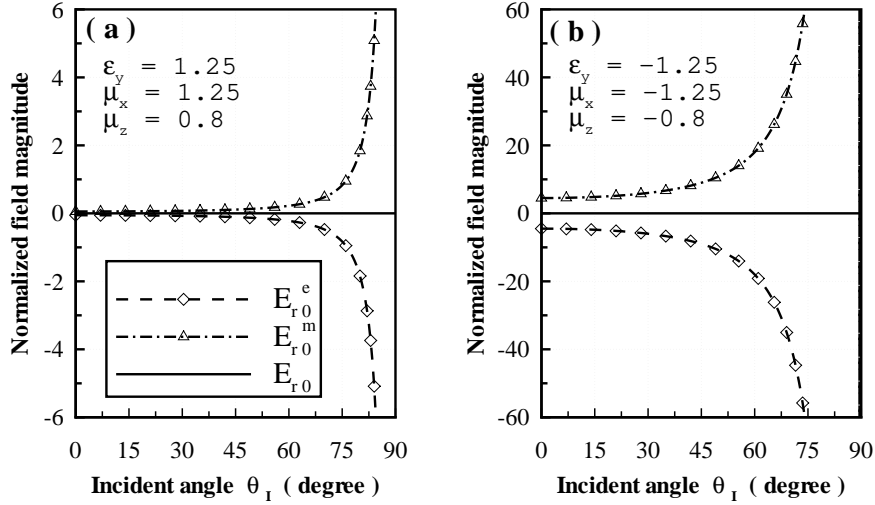}
\caption{\label{Fig3}Radiated electric field amplitudes for a
E-polarized wave incident from the vacuum into anisotropic
dielectric-magnetic material: (a) $\varepsilon_y=1.25$,
$\mu_x=1.25$, $\mu_z=0.8$; (b) $\varepsilon_y=-1.25$,
$\mu_x=-1.25$, $\mu_z=-0.8$. Note that $E_{r0}$ is always zero,
since the radiated fields of the oscillating electric and magnetic
dipoles cancel each other for any incident angle.}
\end{figure}

Next we want to investigate the omnidirectional total transmission
of H-polarized incident wave. The appearance of the
omnidirectional total transmission for the H-polarized waves is
due to the reverse roles of electric and magnetic response.
Evidently, interchanging $\varepsilon$ and $\mu$ in
Eq.~(\ref{EO}), we can can get
\begin{equation}
\mu_0 (\varepsilon_0^2- \varepsilon_x \varepsilon_z) \sin^2
\theta_I=\varepsilon_z( \varepsilon_0 \mu_y -
\varepsilon_x\mu_0).\label{HO}
\end{equation}
Analogously, the conditions for omnidirectional total transmission
can be obtained as
\begin{equation}
\frac{\varepsilon_0}{\mu_0}=\frac{
\varepsilon_x}{\mu_y},~~~\varepsilon_0^2 =\varepsilon_x
\varepsilon_z.\label{HOC}
\end{equation}
To obtain a better physical picture of the omnidirectional total
transmission, we plot the radiated fields induced by the electric
and magnetic dipoles in Fig.~\ref{Fig3}. The radiated fields ${\bf
E}_{rad}^e$ and ${\bf E}_{rad}^m$ always have the same magnitude
but exhibit the opposite directions. Hence, the resulting
reflected fields cancel each other for any incident angle.

Comparing Fig.~\ref{Fig3}(a) and Fig.~\ref{Fig3}(b) show that,
although the phenomenon of omnidirectional total transmission is
formally identical (${\bf E}_{r0}\equiv0$), the microscopic origin
are significantly different. It should be emphasized that the
radiated fields ${\bf E}_{rad}^e$ and ${\bf E}_{rad}^m$ in
Fig.~\ref{Fig3}(b) are much larger the counterparts in
Fig.~\ref{Fig3}(a). Hence the molecular optics theory is useful
for uncovering the inherent secret in anisotropic metamaterials.

\section{Conclusion }\label{sec4}
In conclusion, we have investigated the omnidirectional total
transmission of waves incident from vacuum into anisotropic
dielectric-magnetic metamaterials. The omnidirectional total
transmission need not limit at the interface conventional
nonmagnetic anisotropic crystal. If certain conditions are
satisfied, the anisotropic dielectric-magnetic metamaterials
provide more available option to realize the omnidirectional total
transmission. We have shown that the inherent physics underlying
the omnidirectional total transmission are collective operations
of the electric and magnetic responses. We expect many potential
applications based on the total omnidirectional direction
discussed above. They can, for example, be used to construct
refection absent lens, radar-absorbing material and light bending
device.

\begin{acknowledgements}
This work was supported by projects of the National Natural
Science Foundation of China (No. 10125521 and No. 10535010)
and the 973 National Major State Basic Research and
Development of China (G2000077400).
\end{acknowledgements}

\end{document}